# The influence of surface deformation on thermocapillary flow instabilities in low Prandtl melting pools with surfactants

**Amin Ebrahimi[1], Chris R. Kleijn[2], Ian M. Richardson[1]**

[1]Department of Materials Science and Engineering, Delft University of Technology, Mekelweg 2, 2628 CD, Delft, The Netherlands

[2]Department of Chemical Engineering, Delft University of Technology, van der Maasweg 9, 2629 HZ, Delft, The Netherlands

Corresponding author: Amin Ebrahimi (A.Ebrahimi@tudelft.nl)

**Abstract**

Heat and fluid flow in low Prandtl number melting pools during laser processing of materials are sensitive to the prescribed boundary conditions, and the responses are highly nonlinear. Previous studies have shown that fluid flow in melt pools with surfactants can be unstable at high Marangoni numbers. In numerical simulations of molten metal flow in melt pools, surface deformations and its influence on the energy absorbed by the material are often neglected. However, this simplifying assumption may reduce the level of accuracy of numerical predictions with surface deformations. In the present study, we carry out three-dimensional numerical simulations to realise the effects of surface deformations on thermocapillary flow instabilities in laser melting of a metallic alloy with surfactants. Our computational model is based on the finite-volume method and utilises the volume-of-fluid (VOF) method for gas-metal interface tracking. Additionally, we employ a dynamically adjusted heat source model and discuss its influence on numerical predictions of the melt pool behaviour. Our results demonstrate that including free surface deformations in numerical simulations enhances the predicted flow instabilities and, thus, the predicted solid-liquid interface morphologies.

## 1. Introduction

Melt pool formation, the internal molten-metal flow behaviour and the subsequent re-solidification affect the quality of the products manufactured using advanced fusion-based processes such as additive manufacturing and fusion welding [1]. During these processes, a localised energy flux is often employed that melts the material, forms a melt pool and induces large temperature gradients over the melt pool surface. The induced temperature gradients lead to local changes in surface tension of the molten metal that generate thermocapillary forces, which are often recognised as the dominant force driving the molten metal flow [2-4]. When thermocapillary forces are large compared to viscous forces (*i.e.* high Marangoni numbers)—that is typically the case for laser melting applications in welding and additive manufacturing—the fluid flow in melting pools can be very unstable [5-8]. The thermocapillary-driven molten metal flow determines the energy transport in the melt pool and its geometrical evolution, which eventually can affect the chemical and mechanical properties of the parts. A better understanding of the heat and fluid flow in molten metal melt pools is crucial to optimise these advanced fusion-based manufacturing processes and to reduce the number of defects.

Transport phenomena in liquid metal melt pools during laser melting encompass solidification and melting, various types of heat transfer (radiation, convection and conduction), laser-material interactions, surface tension effects, and intensive fluid flow that interact each other and respond nonlinearly to the applied boundary conditions [9]. Real-time experimental measurement of heat and fluid flow in low-Prandtl melting pools (Pr = $O(10^{-1})$) is challenging due to the opacity, high temperature and rapid changes in the system [10] and are mainly limited to two-dimensional interpretations [1]. Thus, numerical simulations with sufficient accuracy are needed to gain a better understanding of the transient heat and fluid flow in low Prandtl melting pools.

Various assumptions have been made in numerical simulations of low-Prandtl melting pools to avoid complexity and to reduce the computational costs such as non-deformable pool surface, 2D-axisymmetric, laminar flow. However, the consequences of making such assumptions on molten metal melt pool behaviour need to be further studied [11, 12]. Making such assumptions may results in the need of tuning the model—through the enhancement of transport coefficient (*i.e.* thermal conductivity and fluid viscosity) or adapting the imposed boundary conditions—that indeed has little physical realism [13-16]. Previous numerical studies on heat and fluid flow in low-Prandtl melting pools with surfactants have shown that the molten metal flow can be very unstable at high Marangoni numbers [6, 7, 15]. However, deformations of the melt pool surface as observed experimentally [5, 17] were not accounted for in those studies. The aim of the present work is to understand the effect of surface deformations on flow instabilities in a liquid metal melting pool with surfactant. Three-dimensional calculations are performed using both laminar and turbulent flow assumptions to predict heat and fluid flow inside the melting pool.

## 2. Problem description

Laser spot melting of a metallic slab is considered in the present work as a representative case according to the experiments performed by Pitscheneder *et al*. [18] and is schematically shown in figure 1. The slab is made of S705 alloy and has a cubic shape with a width ($L$) of $2.4 \times 10^{-2}$ m and a height ($H_m$) of $10^{-2}$ m, and is heated by a defocused laser-beam with a radius of $r_q = 1.4 \times 10^{-3}$ m and a power of $Q = 3850$ W. The laser power density over the surface has a top-hat distribution and only 13% of its power is absorbed by the material surface ($\eta = 0.13$). With this set of parameters, the conduction-mode melting is established in which evaporation is negligible. The base material, which is initially at $T_i = 300$ K, absorbs heat and melts and thermocapillary stresses start to evolve due to the non-uniform temperature distribution over the melt pool surface. Except for the surface tension, the thermophysical properties of the material are assumed to be temperature independent and are reported in table 1. Variations of the surface tension and its temperature gradient as a function of temperature and sulphur content are approximated using the correlation derived by Sahoo *et al*. [19] and are shown in figure 2 for an Fe-S alloy with 150 ppm sulphur content. A gas layer of $H_a = 2 \times 10^{-3}$ m thick is considered above the base material to model free-surface deformations that are caused by the fluid flow in the melt pool.

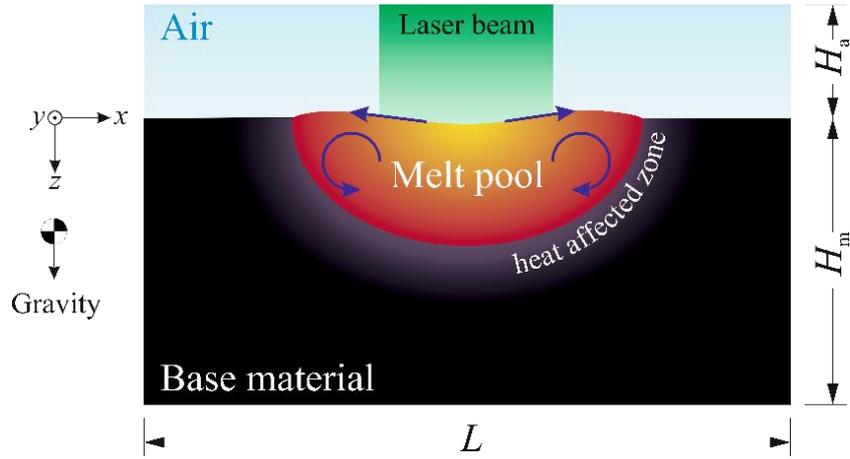

Fig. 1: Schematic of the laser spot melting.

Table 1: Thermophysical properties of the materials employed in the present work.

| Property | Unit | S705 | Air |
| --- | --- | --- | --- |
| Density $\rho$ | kg m$^{-3}$ | 8100 | 1.225 |
| Thermal conductivity $\lambda$ | W m$^{-1}$ K | 22.9 | 0.024 |
| Viscosity $\mu$ | kg m$^{-1}$ s$^{-1}$ | 0.006 | 1.8×10$^{-5}$ |
| Specific heat capacity $c_p$ | J kg$^{-1}$ K$^{-1}$ | 670 | 1006 |
| Latent heat of fusion $L_f$ | J kg$^{-1}$ | 250800 | - |
| Solidus temperature $T_s$ | K | 1610 | - |
| Liquidus temperature $T_l$ | K | 1620 | - |

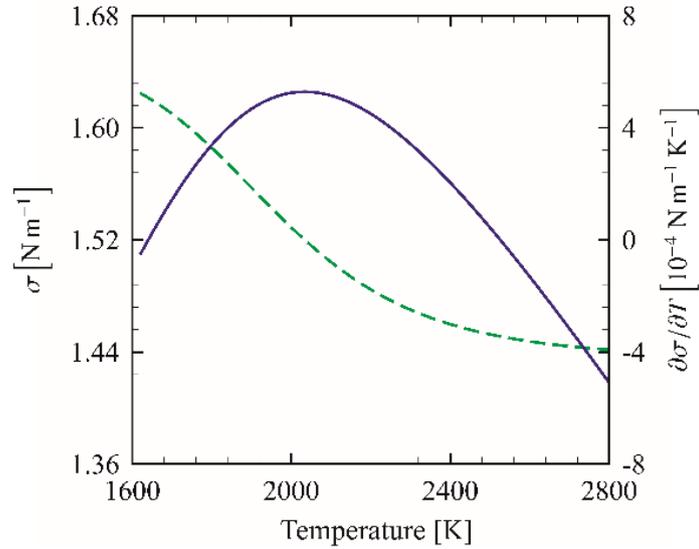

Fig. 2: Variations of surface tension and its temperature gradient as a function of temperature approximated using the Sahoo *et al.* [19] correlation for an Fe-S alloy with 150 ppm sulphur content.

## 3. Governing equations and boundary conditions

To predict heat and fluid flow in the melting pool, a three-dimensional model is employed in which the fluid phases (*i.e.* air and molten metal) are regarded as incompressible and Newtonian fluids. In comparison to thermocapillary-forces that dominate the fluid flow in melting pools during conduction-mode laser spot melting, buoyancy forces are negligible [3], thus, are not accounted for. According to these assumptions, the governing equations for mass, momentum and energy conservations are cast as follows:

$$\frac{\partial \rho}{\partial t} + \nabla \cdot (\rho \mathbf{V}) = 0, \tag{1}$$

$$\frac{\partial (\rho \mathbf{V})}{\partial t} + \nabla \cdot (\rho \mathbf{V}\mathbf{V}) = -\nabla p + \nabla \cdot (\mu \nabla \mathbf{V}) + \mathbf{F}_d, \tag{2}$$

$$\frac{\partial (\rho c_p T)}{\partial t} + \nabla \cdot (\rho \mathbf{V} c_p T) = \nabla \cdot (k \nabla T) + S_e, \tag{3}$$

where, $\rho$ is density, $\mathbf{V}$ fluid velocity vector, $t$ time, $\mu$ viscosity, $p$ pressure, $T$ temperature, $c_p$ specific heat capacity, and $k$ thermal conductivity. $\mathbf{F}_d$ is a momentum sink term that is added to the momentum equation to damp fluid velocities in the mushy region and in solid regions, and is defined as follows:

$$\mathbf{F}_d = \frac{A\mathbf{V}(1-f_l)^2}{f_l^3 + \varepsilon}, \tag{4}$$

where, $A$ is the mushy-zone constant, $\varepsilon$ a small number to avoid division by zero, and $f_l$ the liquid fraction that is a function of temperature, and is defined as

$$f_l = \begin{cases} 0 & T < T_s \\ \dfrac{T - T_s}{T_l - T_s} & T_s \leq T \leq T_l \\ 1 & T > T_l \end{cases} \tag{5}$$

The VOF method [20] was employed to predict the gas-metal interface position for which an additional equation for the transport of a scaler variable $\alpha$ (volume-of-fluid fraction) was solved that reads as follows:

$$\frac{\partial \alpha}{\partial t} + \nabla \cdot (\alpha \mathbf{V}) = 0. \tag{6}$$

The value of $\alpha$ varies between 0 and 1 that represents the gas phase and the metal phase, respectively.

For the cases with a deformable liquid surface, a source term is introduced to the energy equation in the cells located at the metal-gas interface to model the laser heat input and is expressed as

$$S_e = \begin{cases} \dfrac{\lambda \eta Q}{\pi r_q} |\nabla \alpha| \dfrac{2 c_p \rho}{(c_p \rho)_g + (c_p \rho)_m}, & r \leq r_q \\ 0, & r > r_q \end{cases} \qquad (7)$$

where, λ is an adjustment factor to conserve the total heat input [21] at every time-step, and is defined as follows:

$$\lambda = \dfrac{\eta Q}{\iiint_V S_e \mathrm{d}V}. \qquad (8)$$

Atmospheric pressure is applied to the outer boundaries of the gas layer. Since the outer boundaries of the solid regions remain solid in the course of calculations, they are modelled as no-slip walls and heat losses from those boundaries are neglected. For the cases with a non-deformable liquid surface, the gas layer is not modelled explicitly, and the thermal and shear stress boundary conditions are imposed on the material surface that are defined, respectively, as

$$k \dfrac{\partial T}{\partial \mathbf{z}} = \dfrac{\eta Q}{\pi r_q^2}, \quad r \leq r_q \qquad (9)$$

and

$$-\mu \dfrac{\partial \mathbf{V}_t}{\partial \mathbf{z}} = \dfrac{\mathrm{d}\sigma}{\mathrm{d}T} \dfrac{\partial T}{\partial \boldsymbol{\tau}}, \qquad (10)$$

where, **τ** is the tangential vector and $\mathbf{V}_t$ the tangential velocity vector.

## 4. Numerical procedure

A mesh containing $8.6 \times 10^5$ non-uniform cubic cells is used to discretise the computational domain with a minimum cell spacing of $3.5 \times 10^{-7}$ m in the weld pool centre. The cell sizes gradually increase towards the outer boundaries of the computational domain and reach a maximum cell spacing of $2.5 \times 10^{-4}$ m. The model is developed within a commercial solver, Ansys Fluent. The central differencing scheme is used to discretise the convection and diffusion terms of the governing equations. The PRESTO scheme [22] is employed for the pressure interpolation, and the pressure-velocity coupling is handled by the PISO algorithm. [23] The advection of the gas-metal interface is spatially discretised using an explicit compressive VOF scheme [24]. The time derivative is discretised with a first-order implicit scheme. A time-step size of $10^{-5}$ s is chosen to obtain a Courant number (Co = |**V**|Δ*t*/Δ*x*) less than 0.3. Each simulation is executed on 40 cores (Intel Xenon E5-2630) of a computer cluster. Large Eddy Simulations based on the dynamic Smagorinsky-Lilly model [25] is utilised to model the turbulence. Scaled residuals of the governing equations of less than $10^{-7}$ is used as the convergence criterion.

## 5. Results and discussion

### 5.1. Validation of the numerical model

Figure 3 shows a comparison of the melt pool shape predicted using the present model and the experimental observation of Pitscheneder *et al.* [18], which indicate a reasonable agreement. However, it should be noted that the thermal conductivity and viscosity of the liquid metal are artificially increased by a factor of 7 in these simulations (known as "enhancement factor"), as suggested by Pitscheneder *et al.* [18] to achieve an agreement between the numerically precited melt pool shape and the experimentally observed post-solidification macrographs. The heat and fluid flow structure differ notably if an enhancement factor is not employed [14]. The need of such unphysical enhancement factor is mainly due to the neglect of relevant physics in the numerical simulations [13, 14]. Since the focus of the present study is to understand the heat and fluid flow in the melting pools and not the post-solidification melt pools shape, the results presented in the next sections are obtained without using an enhancement factor.

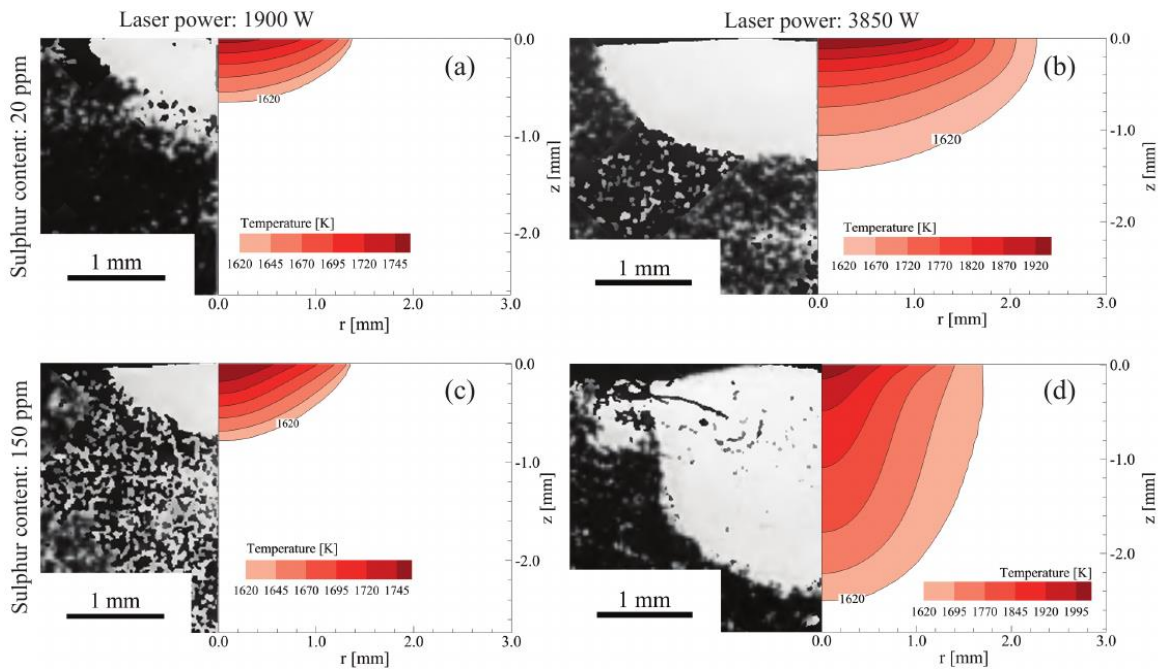

Fig. 3: Melt-pool shapes obtained from the present numerical simulation for different laser power inputs and different sulphur contents after 5s of heating and the experimentally determined post-solidification melt-pool shapes reported by Pitscheneder *et al.* [18].

### 5.2. The influence of free surface deformations

A melt pool forms in the material after a certain time of being exposed to the heat generated by a laser beam. Figure 4 shows the temperature distribution and the velocity vectors over the melt pool surface at *t* = 4s. The non-uniform temperature distribution over the melt pool surface changes the surface tension locally that results in the generation of thermocapillary stresses over the melt pool surface, which drive the fluid flow inside the melt pool. The flow moves from the melt pool centre towards its rim and meets an inward flow from the rim because of the change in the sign of surface tension temperature coefficient (see figure 2). This fluid flow is found to be unstable and asymmetric, which is mainly because of the unstable interactions between the opposing fluid flows at the melt pools surface, and the changes in the melt pool shape [6, 7]. The predicted temperatures over the melt pool surface are 7-16% lower when the liquid surface is assumed to remain flat compared to the case with a deformable surface. Additionally, the predicted fluid velocities are roughly 25% lower for the case with a non-deformable surface, which is due to the lower temperature gradients generated over the pool surface. The higher temperature gradients predicted in the case of deformable surface lead to higher thermocapillary stresses that are sensitive to spatial disturbances resulting in destabilising effects. Variations of temperature at the melt pool centre are shown in figure 5, which demonstrate fluctuations with larger amplitudes when free surface deformations are taken into consideration. When the melt pool surface is assumed to remain flan and non-deformable, temperature fluctuations of the order of 20 K are obtained, which are smaller than those obtained from the simulations with a deformable free surface.

The value of the Péclet number, which indicates the relative importance of advection to diffusion, is significantly larger than one (Pe = $O(10^2)$). Thus, the flow pattern in the melt pool affects the energy transport and the solid-liquid phase transformations. The predicted melt pool shape at *t* = 4s for the case with a deformable surface and the case with a non-deformable surface is shown in figure 6. The enhancement of convection in the melt pool when surface deformations are accounted for results in a shallow bowl-shaped melt pool, which indeed differs from the doughnut-shape pool obtained from the simulations with a non-deformable surface.

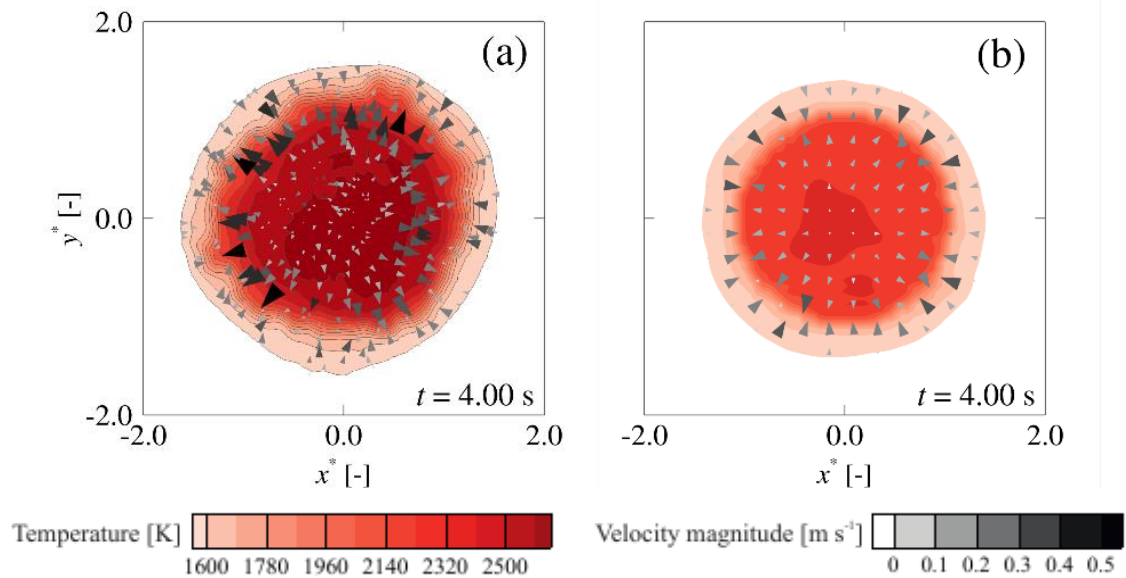

Fig. 4: Temperature distribution and the velocity vectors coloured by their magnitude over the melt pool surface at t = 4s. (a) the case with deformable surface and (b) the case with a non-deformable surface. The laser beam radius is used to non-dimensionalised the coordinates.

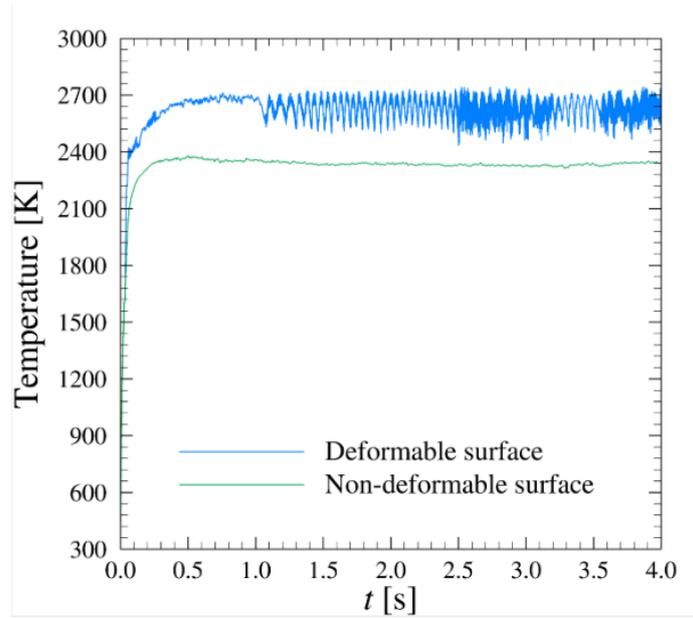

Fig. 5: Temperature fluctuations at the melt pools centre for the cases with deformable and non-deformable surfaces.

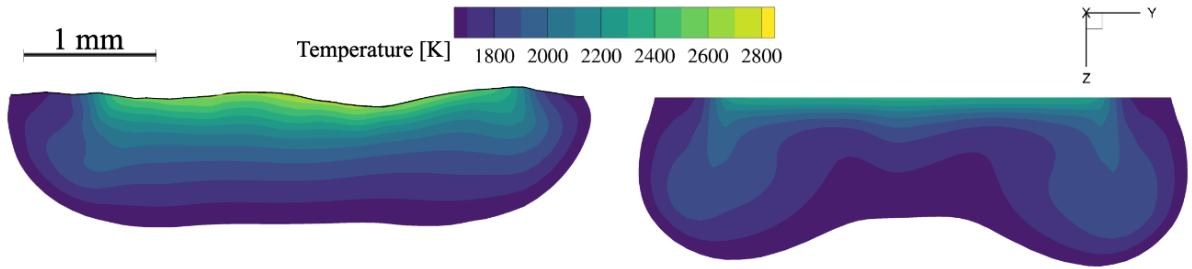

Fig. 6: Predicted melt pool shape at *t* = 4s using a deformable surface (left) and a non-deformable surface (right).

### 5.3. The influence of incorporating the turbulence model

The influence of incorporating the turbulence model in numerical simulations is investigated using Large Eddy Simulations based on the dynamic Smagorinsky-Lilly model [25]. Temperature signals recorded at the melt pool centre from different simulations are presented in figure 7. The incorporation of the turbulence model results in the reduction of the amplitude of temperature fluctuations, which is mainly due to the enhancement of the effective viscosity and thermal conductivity in the melt pool. However, the difference between the predicted temperature fluctuations from simulations with and without the turbulence model is not significant for the case studied in the present work.

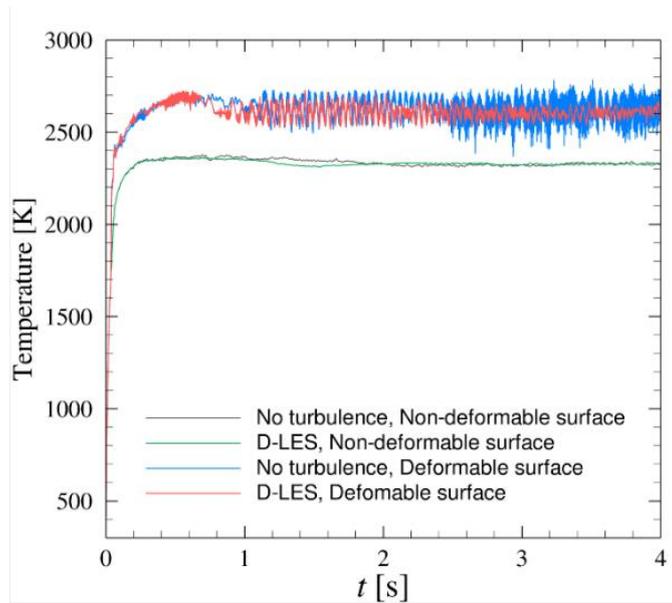

Fig. 7: Temperature fluctuations predicted at the melt pools centre for different cases studied in the present study.

## 6. Conclusions

Heat and fluid flow in low-Prandtl melting pools was studied numerically to understand the effect of surface deformations on thermocapillary flow instabilities. For the case considered in the present work, self-excited thermocapillary flow instabilities were observed that results in three-dimensional unstable fluid flow inside the melt pool. Free surface deformations can affect the heat and fluid flow pattern in the molten metal melt pool, even if they are small compared to the pool size. When surface deformations are not accounted for in simulations, the predicted thermal and velocity fields are different resulting in a different pool shape. Additionally, compared to the influence of free surface deformations on melt pool behaviour, the incorporation of the turbulence model has a negligible influence on numerical predictions for the case studied in the present work.

**Acknowledgements**

This research was carried out under project numbers F31.7.13504 in the framework of the Partnership Program of the Materials innovation institute M2i (www.m2i.nl) and the Foundation for Fundamental Research on Matter (FOM) (www.fom.nl), which is part of the Netherlands Organisation for Scientific Research (www.nwo.nl). The authors would like to thank the industrial partner in this project "Allseas Engineering B.V." for the financial support.